\documentclass{osa-article}

\journal{oe}
\usepackage[utf8]{inputenc}
\usepackage[english]{babel}
\usepackage{graphicx}% Include figure files
\usepackage{dcolumn}% Align table columns on decimal point
\usepackage{bm}% bold math
\usepackage{amsmath}
\usepackage{color}
\usepackage{comment}
\usepackage{array}
\usepackage{chemmacros}
\usepackage{textgreek}
\usepackage{siunitx}
\usepackage{comment}
\newcolumntype{P}[1]{>{\centering\arraybackslash}p{#1}}
\usepackage{soul}
\usepackage{xcolor}

\usepackage{hyperref}
\hypersetup{
     colorlinks   = true,
     linkcolor    = blue,
     citecolor    = blue,
     urlcolor = blue
}

\begin{document}

%%%%%%%%%%%%%%%%%%% title %%%%%%%%%%%%%%%%
\title{Target-wavelength-trimmed second harmonic generation with gallium phosphide-on-nitride ring resonators}

\author{Lillian Thiel,\authormark{1,*} Alan D. Logan,\authormark{1} Srivatsa Chakravarthi,\authormark{1} Shivangi Shree,\authormark{2} Karine~Hestroffer,$^{3}$ Fariba~Hatami,$^{3}$ and Kai-Mei C. Fu\authormark{1,2,*}}

\address{\authormark{1}Department of Electrical and Computing Engineering, University of Washington, Seattle WA 98195\\
\authormark{2}Department of Physics, University of Washington, Seattle WA 98195\\
\authormark{3} Department of Physics, Humboldt-Universitat zu Berlin, 12489 Berlin, Germany}

\email{\authormark{*}lhthiel@uw.edu} 
\email{\authormark{*}kaimeifu@uw.edu} 

%%%%%%%%%%%%%%%%%%% abstract %%%%%%%%%%%%%%%
\begin{abstract}

We demonstrate post-fabrication target-wavelength trimming with a gallium phosphide on silicon nitride integrated photonic platform using controlled electron-beam exposure of hydrogen silsesquioxane cladding. A linear relationship between the electron-beam exposure dose and resonant wavelength red-shift enables deterministic, individual trimming of multiple devices on the same chip to within 30\,pm of a single target wavelength. Second harmonic generation from telecom to near infrared at a target wavelength is shown in multiple devices with quality factors on the order of $10^4$. Post-fabrication tuning is an essential tool for targeted wavelength  applications including quantum frequency conversion.
\end{abstract}

%%%%%%%%%%%%%%%%%%%%%%%%%%  body  %%%%%%%%%%%%%%%%%%%%%%%%%%
\section{Introduction}

Nanophotonic devices have transformed frequency conversion processes, enabling conversion efficiencies orders of magnitude higher than in bulk materials by confining light and controlling mode dispersion. 
To fully leverage the potential of nanophotonic platforms for frequency conversion, multi-resonant devices are desirable; for example, in a $\chi^{(2)}$ nonlinear media, doubly resonant devices for second-harmonic generation (SHG) and parametric down-conversion (PDC) and triply resonant devices for difference- and sum-frequency generation. 
Moreover, applications are emerging in which an absolute frequency of one or more interacting fields is required. Examples of this include PDC entangled photon sources for optical quantum computing~\cite{kwiat1999ultrabright,mosley2008heralded,yao2012observation,wang2021integrated} and frequency conversion of single visible or NIR photons, emitted by ions or solid-state defects, for long-distance fiber transmission~\cite{siverns2019neutral,lu2019chip,kumar1990quantum,bock2018high,zaske2012visible}. Toward this latter application, multi-resonant frequency conversion of photons to/from the telecom band has been demonstrated in nanophotonic structures fabricated from non-linear materials such as gallium phosphide (GaP) ~\cite{logan_2018}, aluminum nitride~\cite{bruch201817}, lithium niobate~\cite{rao2019actively,chen2019ultra} and silicon nitride~\cite{liu2013electromagnetically,lu2019efficient}. However, tight fabrication tolerances for such devices typically cause resonant wavelengths to differ significantly from their designed values, presenting a challenge for practical device implementation. %Hence, the scalable implementation of frequency conversion devices will require robust post-fabrication control of device parameters. 
Here we introduce a promising technique for post-fabrication nanoscale control of photonic resonances to address this challenge via electron-beam modification of hydrogen silsesquioxane (HSQ) cladding.  

Generally, photonic resonators can be tuned by altering the physical dimensions or optical properties of the structure. For a ring resonator, tuning variables include ring width and height, resonator refractive index, and cladding refractive index. Resonances may be either tuned via processes that require constant active control ({\it e.g.} temperature~\cite{Dong2009, Koehler2018}, electric field~\cite{Jung2014, Shen2010}) or trimmed via processes that permanently change device parameters ({\it e.g.} etching~\cite{Hennessy2005, Lu2019, atabaki2013accurate}, introducing strain~\cite{milosevic2018ion, schrauwen2008trimming}, modification of cladding material~\cite{spector2016localized,biryukova2020trimming,zhou2009athermalizing,prorok2012trimming}). 
Resonant wavelength modification of ring resonators for frequency conversion processes presents a particularly complex challenge due to the simultaneous involvement of multiple wavelengths. Any tuning or trimming process will cause a shift in both absolute resonant wavelength and relative spacing between the two or more required resonances. Thus, realizing multiple independent tuning mechanisms and/or developing techniques which offer the spatial resolution to preferentially tune certain resonator modes are particularly interesting. Further, a tuning mechanism that can be readily integrated into a well established fabrication process is preferred.

HSQ is heavily utilized for high-resolution electron-beam lithography~\cite{grigorescu2009resists,manfrinato2019patterning} of photonic devices and is fully compatible with most materials and fabrication processes. Prior studies have investigated the utility of HSQ for resonance tuning using thermal annealing~\cite{yang2002structures} to convert the cage-like structure of uncured HSQ into a denser cross-linked silica structure, increasing the index of refraction from n\,=\,1.45 to n\,=\,1.51. Similarly, localized laser\cite{biryukova2020trimming} and {\it in situ}\cite{spector2016localized} annealing of HSQ on individual silicon micro-ring resonators has been utilized to shift resonances by up to 2\,nm. Both techniques involve heating the devices to high temperatures, thus limiting integration of III-V non-linear photonic materials such as GaAs and GaP. Additionally, long-term, the spatial resolution afforded by thermal processes will be inadequate for multi-resonance frequency conversion applications. High-energy electron-beam exposure allows high spatial and tuning resolution ~\cite{prorok2012trimming}. In this work we investigate the utility of HSQ electron-beam exposure as a method for high-resolution, post-fabrication, single device trimming, with the potential for mode-selective trimming. Electron-beam exposure of HSQ has been shown to induce cross-linking similar to thermal annealing, while high exposure intensity enables further refractive index tuning (up to $\mathrm{n=1.62}$) via the formation of silicon-rich SiO$_2$~\cite{lee2000planarhsq,choi2008comparative}. Our results indicate that the induced wavelength shift is linearly dependent on the electron-beam exposure dose, with an observed trimming range of $\sim$\,5\,nm for the applied doses. We demonstrate high spatial and tuning resolution, resonance stability, and minimal impact on device quality factor (Q), making HSQ electron-beam trimming a promising candidate for use in quantum frequency conversion applications. Finally, we demonstrate successful target-wavelength second harmonic generation with multiple devices and provide an outlook toward utilizing this technology for simultaneous multiple resonance tuning.

\begin{figure}[t]
\centering
\includegraphics[width=0.9\textwidth]{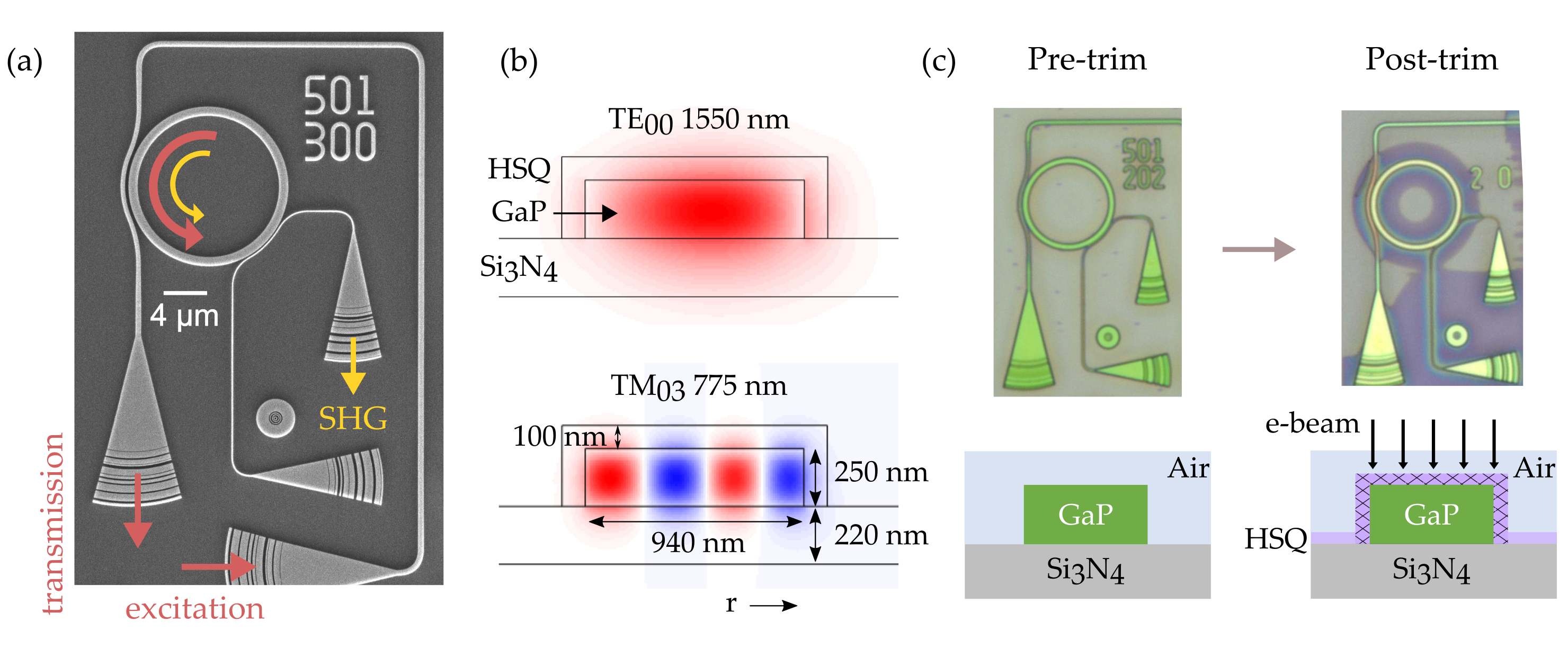}
\caption{(a) SEM image of the fabricated GaP-on-Si$_3$N$_4$ resonator with accompanying grating couplers. The device is excited with telecom band tunable laser (red) and the SHG signal is measured out of the vis/NIR output grating (yellow). (b) Simulated resonator $\vec{E}$-field profiles of the quasi-phase matched telecom and second-harmonic modes. (c) Pre and post trimming optical images of a device. The ring was exposed with a donut pattern. The waveguide and grating structures were utilized for exposure alignment. To reveal the exposed HSQ pattern during the trimming process, the post-trimmed optical image was acquired after developing the exposed HSQ in a 25\% TMAH solution. Note that all optical characterizations were performed before the development.}
\label{fig:devices}
\end{figure}

%--------------------------------------------------------------------
%--------------------------------------------------------------------
\section{Resonator design, fabrication and testing}

%Design
The ring resonators were designed and fabricated in a 250\,nm (100) GaP photonic layer on an epiwafer substrate with 220\,nm LPCVD silicon nitride on 4\,\textmu m silicon oxide on silicon. A ring radius of 7.77\,\textmu m and width of 940\,nm were selected to satisfy a quasi-phase matching condition for a TE$_{00}$ fundamental mode and  TM$_{03}$ second harmonic (SH) mode at $\mathrm{\lambda_0 \approx 1550\,nm}$ and $\mathrm{775\,nm}$, respectively~\cite{logan_2018}. % Actual ring pattern widths vary from 900 to 964 nm, to improve chances that some fabricated devices would be tunable to double resonance
Each ring mode is evanescently coupled to a different input/output waveguide leading to a pair of cross-polarized grating couplers. A device SEM image and schematic are shown in Fig.~\ref{fig:devices}a, with the relevant mode profiles in Fig.~\ref{fig:devices}b. Additional design details can be found in Appendix~\ref{appendix:design}. 

Eigenmode simulations of the ring resonator predict that an increment in refractive index of $\mathrm{\Delta n_{HSQ}=+\,0.1}$ in a 100\,nm thick conformal HSQ cladding will red-shift the resonant wavelength of the fundamental mode by $\sim$\,2.98\,nm and SH mode by $\sim$\,1.07\,nm. With changes in HSQ refractive index as high as 0.22 reported~\cite{lee2000planarhsq,choi2008comparative}, the predicted trimming range is capable of compensating for typical device-to-device fabrication variations.

%Fabrication
To fabricate the structure, the 250\,nm GaP layer was released from a GaP substrate by etching an intermediate $\mathrm{Al_{0.8}Ga_{0.2}P}$ sacrificial layer with dilute HF. The GaP was then transferred to the  Si$_3$N$_4$-on-SiO$_2$/Si substrate using a wet transfer process~\cite{yablonovitch_van_1990, gould_efficient_2016, schmidgall_frequency_2018,logan_2018}. Electron-beam lithography using HSQ resist and subsequent plasma reactive-ion-etching (RIE) of the GaP layer forms the photonic devices. Two chips (A and B) were fabricated with identical device dimensions. After fabrication, a layer of developed HSQ mask remains. This HSQ layer is removed with a vapor-HF etch, enabling the spin-on of a new conformal coating of HSQ for the resonance trimming process. After HSQ spin-on, each device was trimmed by exposing a donut-shaped region extending 1\,\textmu m beyond the inner and outer device radii at a beam voltage of 5\,keV (Fig.~\ref{fig:devices}c). Further details of the fabrication are provided in Appendix ~\ref{appendix:fabrication}. The fabrication process is compatible with diamond substrates for GaP-on-diamond or GaAs-on-diamond photonic circuits integrated with solid-state defects~\cite{gould_efficient_2016, schmidgall_frequency_2018,huang2021hybrid}.

%Testing
The fundamental resonances of individual devices were identified by sweeping a continuous-wave telecom laser (1530 to 1565\,nm) and detecting transmission spectra using a InGaAs photodiode. In the visible band a supercontinuum laser (550 to 1350\,nm) was used for excitation and the transmission spectra is detected with a grating spectrometer. The excitation polarization was set to excite TE (1550\,nm) or TM (775\,nm) modes, as depicted in  Fig.~\ref{fig:devices}a. Scattered excitation light was filtered out of the collection path via polarization and spatial filtering. Chip-A exhibited resonances with Q-factors of 7\,$\pm$\,2.2\,$\times$10$^3$. Chip-B exhibited narrower resonances with Q-factors of 3\,$\pm$\,1.7\,$\times$10$^4$.

%--------------------------------------------------------------------
%--------------------------------------------------------------------
\section{HSQ target-wavelength trimming}

\begin{figure}[t]
\centering
\includegraphics[width=0.9\textwidth]{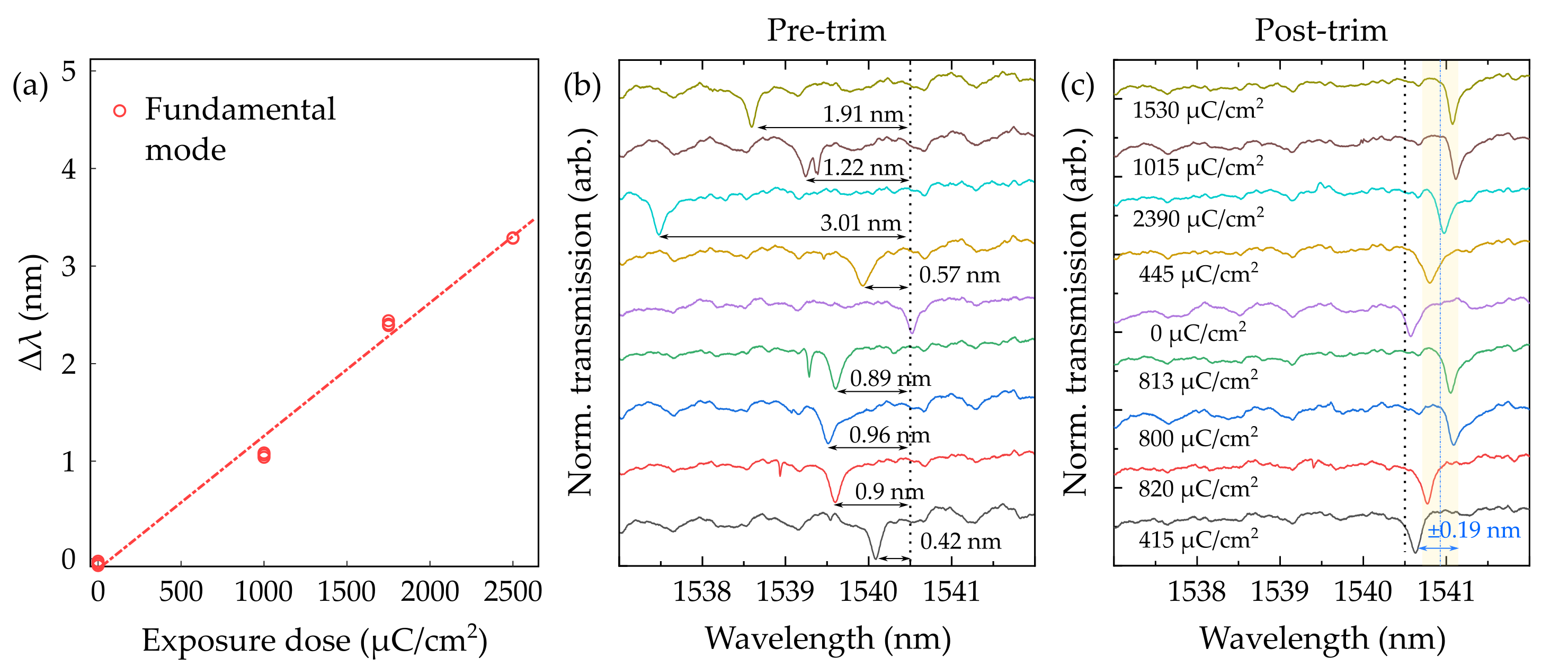}
\caption{Chip-A: (a) Relationship between HSQ exposure dose and observed wavelength shift at the fundamental wavelength. (b), (c) A target trim-test showing the fundamental wavelength resonances for nine devices with identically designed resonator dimensions pre- and post-exposure, respectively. The target wavelength was selected to be 1540.5\,nm, however a systematic overexposure resulted in resonances trimmed to 1540.91$\pm$0.19\,nm.}
\label{fig:trimming_data}
\end{figure}

The first goal is to establish a relationship between exposure dose and resonant wavelength shift to deterministically tune the resonant wavelength of individual devices on-chip. Approximately 24 hours after HSQ spin-on, 13 devices (chip-A) were selected for a dose-test and exposed with electron-beam lithography. Three devices each received doses of 1000, 1750, and 2500\,\textmu C/cm$^2$ respectively and four devices received no exposure. Transmission spectra of one specific fundamental mode taken before and after exposure show a linear relationship between resonance shift and exposure dose (Fig.~\ref{fig:trimming_data}a). On average, the exposed devices exhibit a tuning rate (red-shift) of 1.4\,$\pm$0.1 pm/\textmu C/cm$^{2}$, with the maximum tested dose (2500\,\textmu C/cm$^{2}$) giving an average shift of 3.08\,nm. 

In the next step, a set of nine identically designed devices were identified (Fig.~\ref{fig:trimming_data}b) with initial resonances near $\mathrm{\lambda = 1540.5}$\,nm to trim to mutual resonance. Before trimming, the spread of resonant wavelengths was measured to be $\approx$\,3\,nm. We used the above linear dose-shift relationship to determine the necessary exposure dose for targeted trimming of each device resonance to $\mathrm{\lambda = 1540.5}$\,nm. With a single exposure of each device, all resonances were trimmed to the target wavelength with an offset of 0.41$\pm$0.19\,nm (Fig.~\ref{fig:trimming_data}c). 

Post-tuning analysis (Appendix \ref{appendix:mode}) of the transmission spectra determined that the tuning rate is different for various families of modes supported by the ring. The highest observed shift was 4.92\,nm (Fig.~\ref{fig:appendix1}a, mode-1). The significant mode-sensitivity suggests deterministic inter-modal control could be developed within this platform. Further, we observed an average post-trim resonance drift of 10.6\,pm/day (corresponding to 0.7\% of average trimmed red-shift or 5\% of the average target resonance linewidth) over 12 days. This minor drift followed a non-monotonic trend for all tested devices. The source of the post-exposure drift requires further investigation.

%--------------------------------------------------------------------
%--------------------------------------------------------------------
\section{Target-wavelength-trimmed SHG}

\begin{figure}[t]
\centering
\includegraphics[width=0.9\textwidth]{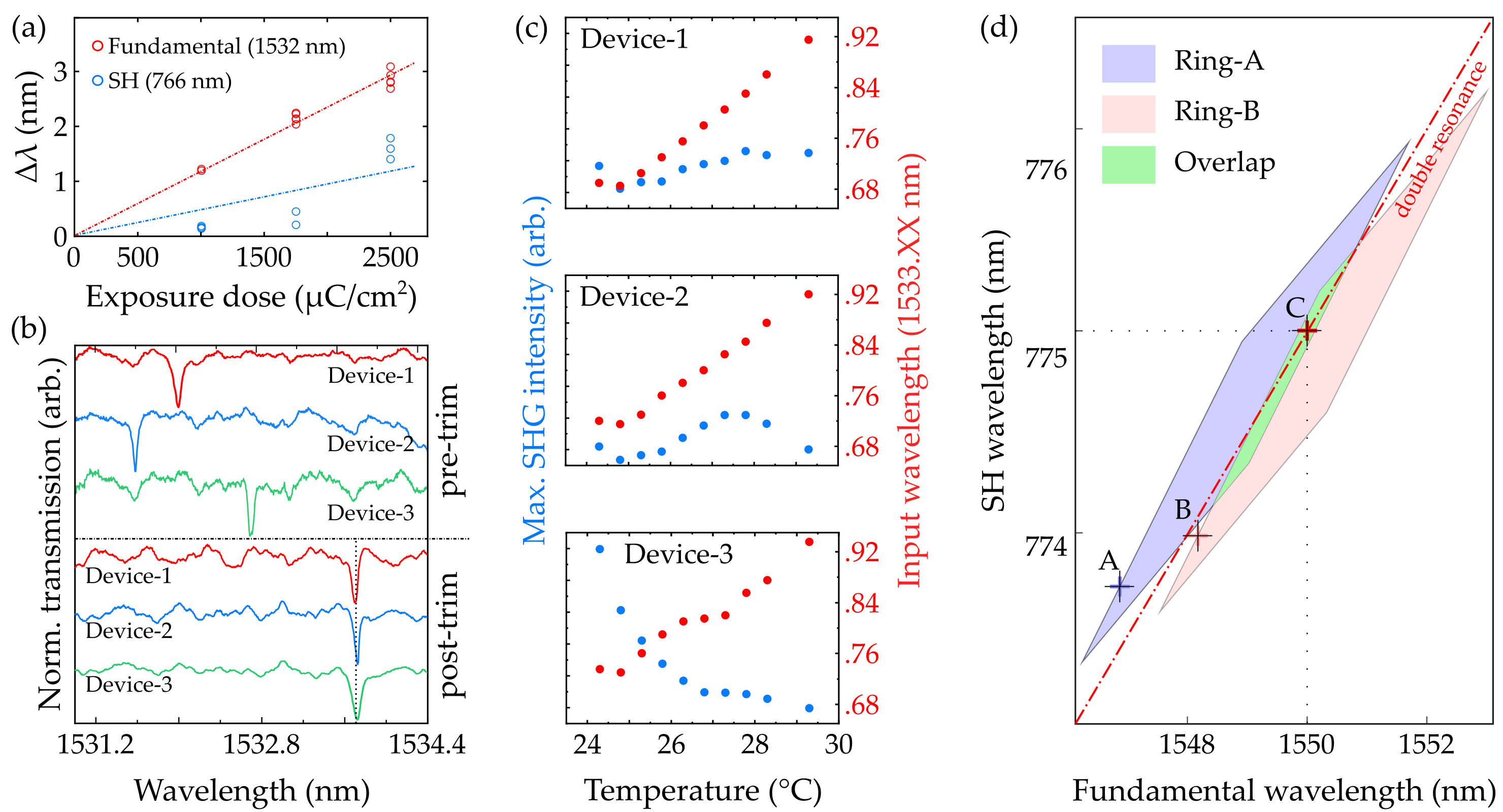}
\caption{Chip-B: (a) Linear dependence of resonant wavelength shifts with the doses for both fundamental and SH (blue) wavelengths with estimated slopes of 1.18 and 0.47 pm/\textmu C/cm$^{2}$ respectively. (b)The dose-test data is utilized to trim the fundamental wavelength resonance of three near-quasi-phase-matched devices.
(c) Peak SHG intensity (blue) and corresponding input (fundamental resonance) wavelength (red) as a function of temperature for each trimmed device. The fundamental and SH resonance red-shift linearly with different rates with increasing temperature. The SHG intensity depends on how close the device is to double resonance (SH resonance at exactly half the input wavelength). (d) A simulated trimming scenario showing two ring resonators with different initial fundamental and SH resonant frequencies (points A and B). Blue and red regions show the range of resonant wavelengths that can be reached using a combination of temperature tuning (10 to 60$^{\circ}$C) and HSQ exposure (0 to 2500\,\textmu C/cm$^2$), with the overlap (green) showing that each ring could be tuned to double resonance (dotted red line) over a range of wavelengths including the design wavelength (point C). Wavelength shift dependence on temperature and HSQ exposure are derived from experimental data for the fundamental and simulations for the SH resonances.}
\label{fig:trimming_data2}
\end{figure}

In a second experiment (chip-B), we expanded the scope of our dose-test to incorporate both the fundamental and SH modes. Using the same dose-test procedure, a linear dose-shift relationship is found for the fundamental mode with a nearly identical tuning rate (1.2\,$\pm$0.1 pm/\textmu C/cm$^{2}$) to the chip-A dose-test. The largest dose (2500\,\textmu C/cm$^2$) corresponds to an average shift of 3.0\,nm. This matches the simulated resonance shift for a cladding index change of $\mathrm{\Delta n_{HSQ} \approx 0.1}$, which is reasonable for the large electron-beam exposure dose~\cite{lee2000planarhsq}. A linear dose-shift relationship (0.5\,$\pm$0.2 pm/\textmu C/cm$^{2}$) is also established for the SH mode (Fig.~\ref{fig:trimming_data2}a). Due to the highly multi-mode nature of the ring at 775\,nm, difficulty in consistently identifying radial modes prevented us from accurately determining the mode-specific trimming rate of the targeted SH mode. Thus, we focus on HSQ target trimming of the fundamental mode to demonstrate target-wavelength-trimmed SHG in multiple devices. 

We selected three nominally identical devices with telecom resonances that exhibited strong SH conversion. The initial spread of resonances was 1.11\,nm. Each device was exposed to a calibrated HSQ electron-beam dose for trimming to a target wavelength of $\mathrm{\lambda = 1533.75}$\,nm. All three devices were trimmed to within $\mathrm{\approx30\,pm}$ of the target wavelength (Fig.~\ref{fig:trimming_data2}b). The strong temperature dependence of the SHG signal (Fig.~\ref{fig:trimming_data2}c), indicates that the devices were close to quasi-phase matched double resonance after the trimming process. The SH and fundamental resonances red-shift at different rates with increasing temperature, causing the SHG conversion efficiency to peak as the SH resonance tunes into and out of double resonance with the fundamental. All three devices gain some double-resonant SHG enhancement at the target wavelength (1533.75\,nm), but due to remaining variations in the SH resonant wavelength after the HSQ exposure trimming, each resonator reaches its maximum SHG efficiency at a slightly different temperature and wavelength.

SHG efficiency is also very sensitive to the Q-factors of the resonant modes. The narrow resonances observed on chip-B enable us to investigate the effect of trimming on the fundamental mode Q-factor. Initially the average Q-factor for 15 devices was 3\,$\pm$1.7$\times$10$^{4}$. The Q-factor is observed to both decrease and increase with the average Q decreasing by 5.1\,$\pm$6.1$\times$10$^{3}$ after exposure. It is likely that the majority of this variation in Q can be attributed to exposure of the coupling region, altering the coupling-Q as described in Appendix \ref{appendix:design}. An exposure pattern that avoids the coupling region could maintain the device Q through the trimming process. Further, selective exposure of the coupling regions could be utilized to tune devices to the critical-coupling regime in addition to resonant wavelength trimming.

%--------------------------------------------------------------------
%--------------------------------------------------------------------
\section{Outlook and Conclusion}

In this work we demonstrated that HSQ exposure is a promising post-fabrication trimming method for photonic resonators. With a simple calibration process, we target-wavelength trim a large set of identically designed devices with typical post-fabrication variations. HSQ trimming will be an important tool in the larger effort towards target-wavelength frequency conversion. For example, combined with temperature tuning, HSQ trimming should enable dual target-wavelength trimming for doubly resonant quasi-phase matched frequency conversion processes. The devices in this work that were trimmed to correct variance at the telecom wavelength at room temperature still had some variance in the SH resonance, resulting in maximum SHG conversion at different temperatures and wavelengths. By taking the effect of temperature tuning into account when selecting exposure doses, Fig.~\ref{fig:trimming_data2}d illustrates how multiple resonators could be trimmed to provide their maximum SHG conversion efficiency at the same target wavelength. Two simulated devices (A, B) with fundamental wavelengths differing by 1.3\,nm at room temperature and $\mathrm{\sim5.5\,nm}$ at their optimal SHG temperatures can be target wavelength tuned within the green dose-temperature area. A third independent tuning mechanism will be required to achieve optimal efficiency for a specific sum- or difference-frequency conversion process needed for quantum frequency conversion applications.

The attractive properties of HSQ exposure trimming include tuning range, resolution and precision, low post-trim drift and minimal effect on quality factor. We expect that additional process development will further improve the accuracy and precision. Better mode identification will reduce uncertainty in mode-specific trimming rates enabling the implementation of multiple tuning mechanisms. Improved understanding of sources of drift will allow us to design for or mitigate post-trim resonant shifts. The spatial resolution afforded by electron-beam lithography for HSQ trimming creates potential for applications beyond resonance shifting. Selective exposure of coupling regions could allow post-fabrication fine tuning of coupling Q. Other works have shown that periodic nanometer-scale variations along the ring inner radius can target specific modes for mode-splitting~\cite{Lu2020}. Electron-beam lithography-based trimming makes target mode trimming and splitting a realistic opportunity. We have shown HSQ exposure to be a simple and effective post fabrication trimming technique. With further development it stands to become a process widely utilized by the photonics community.

%--------------------------------------------------------------------
%--------------------------------------------------------------------
%--------------------------------------------------------------------
%--------------------------------------------------------------------
%To fix section and figure numbering for appendix
\renewcommand{\thefigure}{A\arabic{figure}}
\setcounter{figure}{0}
\renewcommand{\thesection}{A\arabic{section}}
\setcounter{section}{0}
\vspace{2em}

\section*{Appendix}

\section{Device design and simulation details}
\label{appendix:design}
The ring resonators were designed using bent-waveguide eigenmode simulations in Lumerical MODE. Using ring radius $r$ and width $w$ as design variables, the resonator was optimized to maximize nonlinear overlap $\beta$ while satisfying the mode polarization and quasi-phase matching (QPM) requirements imposed by the zincblende crystal symmetry and (100)-normal orientation of the GaP photonic layer \cite{logan_2018,bi2012high,ref:boyd2008nlo}. At each design point, the target modes ($\mathrm{TE_{00}}$ at $\mathrm{\lambda_1\,=\,1550\,nm}$ and $\mathrm{TM_{03}}$ at $\mathrm{\lambda_2\,=\,775\,nm}$) were simulated in a ring cross section. The nonlinear overlap $\beta$ was calculated from the mode profiles, along with fractional azimuthal mode numbers $\mathrm{(m_i\,=\,2\pi \cdot r \cdot n_{eff,i}/\lambda_i)}$  for each mode, which are used to check the QPM condition $\mathrm{(2m_1\,-\,m_2\,=\,\pm\,2)}$. Integer azimuthal mode numbers at the simulated wavelengths are preferred, but not required; a design with non-integer $m_i$ is expected to have QPM modes that are near double resonance at wavelengths that are not exactly 1550\,nm and 775\,nm in ambient conditions, which is acceptable if within the tolerance of our tuning capabilities. Based on previous experiments in similar material platforms (GaP-on-Diamond, GaP-on-oxide), intrinsic quality factor is expected to be dominated by fabrication-related factors such as sidewall roughness, so the simulated radiative quality factor was not considered for the design process. 

The ring resonator coupling regions were designed to provide specific coupling quality factors using a supermode analysis method.
For a range of coupling waveguide widths and ring-to-waveguide distances, the target ring and waveguide modes were simulated with each structure alone, and then all guided modes were simulated for the combined structure.
Mode overlap between each coupling region mode and the ring/waveguide modes were calculated and used to numerically find the total power transfer from the ring mode to waveguide mode as a function of coupling region length.
The coupling quality factor $Q_c$ was derived from the simulated free spectral range $\Delta \lambda_{FSR}$ of the resonator mode and the field coupling coefficient $\kappa$ from a single pass through the coupling region: 
\begin{equation}
Q_c = \frac{2 \pi \lambda}{\left| \kappa^2 \right| \Delta \lambda_{FSR}}.    
\end{equation}
To provide a consistent interface to the rest of the coupling photonic circuit, a constant waveguide width and propagation length was selected for each ring mode's coupling region, with only the separation distance varied to set $\mathrm{Q_c}$.

Coupling quality factor depends significantly on how well the HSQ fills the space between the ring and the waveguide, as well as HSQ exposure dose; the coupling regions were simulated with a 100\,nm conformal HSQ layer.
The telecom ring mode is coupled to a 440\,nm wide waveguide wrapped around a $\mathrm{45^{\circ}}$ arc of the ring, and the SH mode is coupled to a 120\,nm waveguide with a $\mathrm{30^{\circ}}$ wrap.
For the telecom mode, a ring-to-waveguide gap distance of 240\,nm gives $\mathrm{Q_c \approx 24000-11000}$ with increasing ebeam exposure, and a distance of 360\,nm gives $\mathrm{Q_c \approx 90000-55000}$.
For the SH mode, a gap distance of 120\,nm gives $\mathrm{Q_c \approx 4500-5500}$, and 160\,nm gives $\mathrm{Q_c \approx 11000-13500}$.
 
The grating couplers were designed to provide coupling to free space with minimal footprint and low back reflections.
For each wavelength band and polarization, an aperiodic grating coupler design was found using a sampled hill climbing optimization algorithm with a simplified model of grating behavior as an objective function. The performance of the final design was verified with a 3D FDTD simulation.
To suppress on-chip back reflections, which cause Fabry-Perot interference patterns in transmission spectra, the grating designs were implemented as elliptically-shaped focusing grating couplers.
The grating notches are shaped as sections of progressively larger ellipses with one focus centered on the end of the waveguide.
Back reflections from the waveguide are directed toward the second focus of the ellipse instead of going back into the waveguide mode~\cite{vermeulen2012elliptical}.
The design variables used for the elliptical grating were: eccentricity, the angle between waveguide and major axis, the arc angle of the elliptical section, and major axis length of first grating notch.
Designs were evaluated with 3D FDTD simulations to find a combination of parameters that sufficiently suppresses back reflection in the the same mode while maintaining a reasonable spot shape for light scattered into free space.

\section{Fabrication details}
\label{appendix:fabrication}

The GaP material used for the photonics layer was grown by molecular beam epitaxy on a 300\,nm $\mathrm{Al_{0.8}Ga_{0.2}P}$ sacrificial layer on a GaP substrate. This epitaxial GaP layer was released in 2\% HF solution, followed by a H$_2$O rinse and wet transfer to the nitride substrate. To improve GaP adhesion, prior to transfer, a 5\,nm layer of SiO$_2$ was thermally evaporated onto the substrate, which was then treated with hexamethyldisilizane vapor. The transferred GaP was allowed to dry overnight on a hotplate at 80\,$^\circ$C. Electron beam lithography (JEOL-6300, 100\,kV, 1\,nA beam current) was utilized to pattern the designed photonic structures on a thin (150\,nm) HSQ resist layer on top of the GaP-on-nitride stack. Multiple arrays of devices were patterned with the resonator and coupling region dimensions varying across the pattern to compensate for fabrication variations. A reactive-ion etch (RIE) step (3.0\,mTorr, 45\,W RF, 60\,W ICP, 235\,V DC-bias, 1.0/6.0/3.0\,sccm Cl$_{2}$/Ar/N$_2$ flow) was used to etch the GaP layer. The residual HSQ mask had to be removed. Initial attempts at HSQ removal using a dilute HF (0.5\%) dip revealed low adhesion of the GaP devices to the substrate: stiction from the wet process damaged the waveguide coupling regions. Thus, a HF vapor-etch process was investigated. After the successful HSQ vapor-etch, the chip was heated to 200\,$^\circ$C for 5\,min to remove etch byproducts. This provided a clean surface for a $\sim$150\,nm layer of HSQ electron-beam resist that was spun-on and exposed for device trimming. The thickness of this HSQ layer is expected to shrink by up to 33\% due to e-beam exposure~\cite{lee2000planarhsq}.

\section{Post-tuning analysis}
%\section{Mode dependence of HSQ tuning rate }
\label{appendix:mode}

\begin{figure}[t]
\centering
\includegraphics[width=0.9\textwidth]{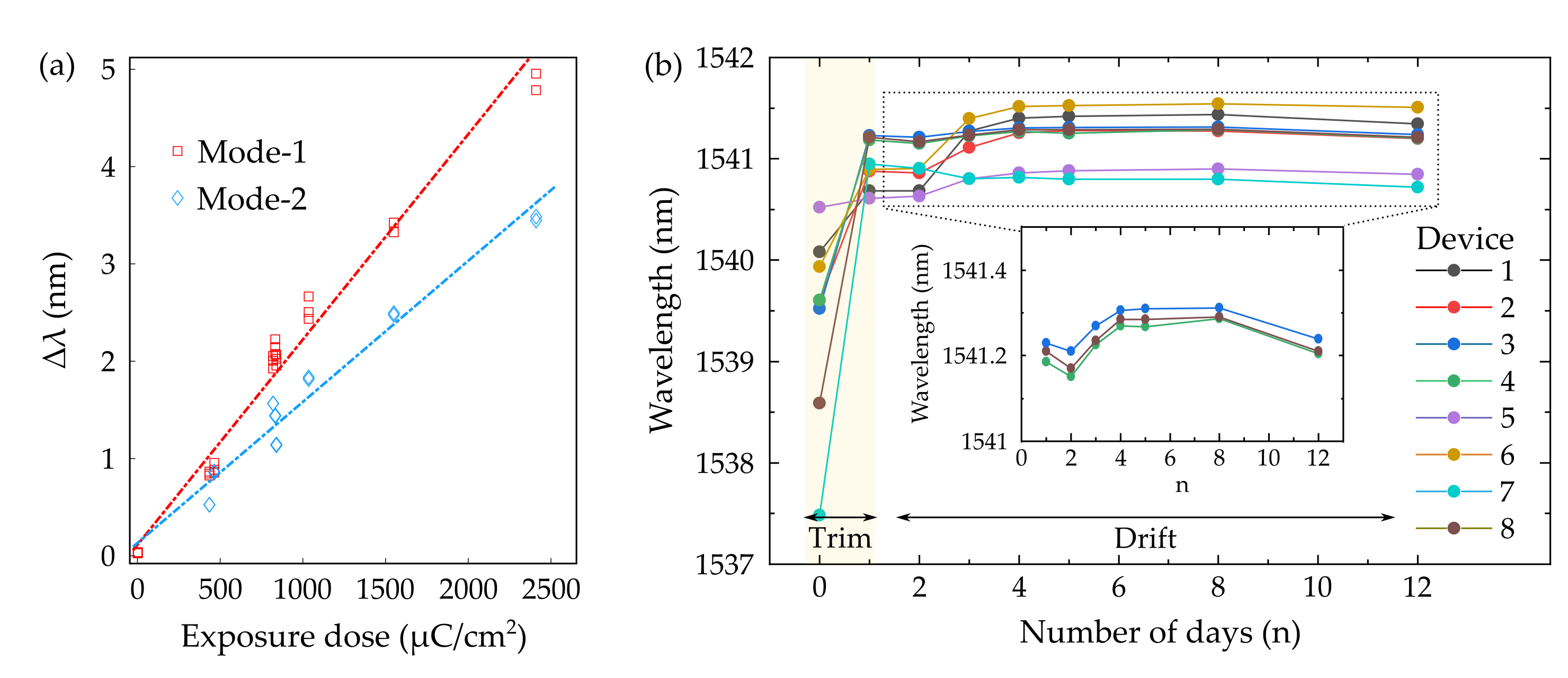}
\caption{(a) Resonance shift as a function of HSQ exposure dose shows a linear relation for two different modes: Mode-1 (red) and Mode-2 (blue) at telecom wavelength range. The slope is different for the two modes indicating that a different dose would be required for the same shift for a specific mode. (b) Temporal stability of the trimming process for a subset of eight devices. Day-0 data shows initial resonance wavelengths of all devices without HSQ exposure. Device 5 did not receive any e-beam exposure. The large shift at day-1 is observed due to coarse e-beam trimming. Devices 1, 2, 6 and 7 were re-exposed for fine target wavelength trimming. 
}
\label{fig:appendix1}
\end{figure}

The dose-shift relationship is dependent on the overlap of the target mode with the surrounding cladding. Analysis of the chip-A trim-test data (Fig.~\ref{fig:appendix1}a) reveals that two families of modes were observed in the transmission spectra (1530 to 1565\,nm) with significantly different tuning rates. Mode-1 (FSR\,=\,13.61\,$\pm$0.14\,nm) exhibits a sensitivity of 2.11\,pm/\textmu C/cm$^2$, and mode-2 (FSR\,=\,12.91\,$\pm$0.05\,nm) exhibits a lower sensitivity of 1.45\,pm/\textmu C/cm$^2$. From simulations, the fundamental TE$_{00}$ mode is expected to have an FSR$\approx$14.7\,nm. Given the smaller observed FSR, we suspect both these families of resonances are higher order hybrid TE$_{01}$/TM$_{00}$ modes. Simulations show these hybrid modes tune about twice as fast as the fundamental mode, with a change in HSQ cladding index of 0.1 shifting the wavelength by about 6.25\,nm and 7.32\,nm respectively instead of 2.98\,nm.

%\section{Temporal stability of the trimmed resonances}
\label{appendix:stability}

To monitor the temporal stability of the trimming process, after HSQ exposure, room temperature transmission spectra of eight identically designed devices were periodically collected over a 12 day period. Devices 1, 2, 6 and 7 were re-exposed for fine target wavelength trimming. Therefore, we only included the remaining devices for detailed analysis of the resonance shift. One possible reason for the observed shift could be the change in refractive index of the cladding over time (possibly from uptake of ambient moisture by the exposed HSQ~\cite{biryukova2020trimming, spector2016localized}) and further investigation is required for identification of the source of the drift. 

%%%%%%%%%%%%%%%% Journal back matter %%%%%%%%%%%%%%%%%%%
\bigskip
\bigskip

\begin{backmatter}

\bmsection{Funding}

\bmsection{Acknowledgments}
This material is based upon work supported by the National Science Foundation under Grants EFMA-1640986 (photonic design and fabrication) and U.S. Department of Energy, Office of Science, National Quantum Information Science Research Centers, Co-design Center for Quantum Advantage (C2QA) under contract number DE-SC0012704 (HSQ tuning and testing). The photonic devices were fabricated at the Washington Nanofabrication Facility, a National Nanotechnology Coordinated Infrastructure (NNCI) site at the University of Washington which is supported in part by funds from the National Science Foundation (awards NNCI-1542101, 1337840 and 0335765). 

\bmsection{Disclosures}
The authors declare that there are no conflicts of interest related to this article.

\bmsection{Data Availability Statement}
Data underlying the results presented in this paper are not publicly available at this time but may be obtained from the authors upon reasonable request.

\end{backmatter}

%%%%%%%%%%%%%%%%%%%%%%% References %%%%%%%%%%%%%%%%%%%%%%%%%
\bibliography{Srivatsa, Shivangi}

\end{document}